\documentclass[12pt]{article}
\usepackage{graphics,graphicx,epsfig}
\makeindex
\newcommand{\be}{\begin{equation}}
\newcommand{\ee}{\end{equation}}
\newcommand{\bea}{\begin{eqnarray}}
\newcommand{\eea}{\end{eqnarray}}

 3
 2
 1
 1
 2
\def\btab{\begin{tabular}}  
\def\etab{\end{tabular}}
\def\rfr#1{eq.(\ref{#1})}
\def\rfrs#1#2{eqs.(\ref{#1})-(\ref{#2})}

\def\bar{\begin{array}}
\def\ear{\end{array}}
\def\eqi{\begin{equation}}
\def\eqf{\end{equation}}
\def\mc#1#2{\left[\matrix{#1 \cr #2\cr}\right]}

\def\ga{\gamma}
\def\d{\delta}

\def\m{\mu}

\def\p{\pi}

\def\t{\tau} 

\def\f{\phi}

\def\og{\omega}
 
\def\D{\Delta}

\def\O{\Omega}

\def\vass#1{\left\vert\ #1 \right\vert}
\def\rp#1#2{{#1\over#2}}
\def\ol#1{\overline{#1}}

\def\derp#1#2{\rp{\partial{#1}}{\partial{#2}}}
\def\dert#1#2{\rp{d{#1}}{d{#2}}}

\def\lb#1{\label{#1}}

\begin{document}
\begin{titlepage}
\begin{flushright}
\today\\
\end{flushright}
\vspace{.5cm}
\begin{center}
{\LARGE The impact of tidal errors on the determination\\ of the
Lense-Thirring effect from satellite laser ranging\\}
\vspace{1.0cm}
\quad\\
{Erricos C. Pavlis$^{\dag}$, Lorenzo Iorio$^*$\\
\vspace{1.0cm}
\quad\\
$^{\dag}$ Joint Center for Earth System Technology, JCET/UMBC
   NASA  Goddard Space Flight Center
   Space Geodesy Branch, Code 926
   ESSB Bldg. 33, Rm G213

   Greenbelt, Maryland
   U S A    20771-0001  \\ \vspace{1.0cm}
\quad\\
\vspace{0.3cm}
\quad\\
$^*$ Dipartimento di Fisica dell' Universit\`a di Bari, via Amendola 173,
70126,
Bari, Italy\\ \vspace{1.0cm}
}
\vspace*{1.0cm}

{\bf Abstract\\}  
\end{center}

{\noindent \footnotesize
The general relativistic Lense-Thirring effect can be detected by 
means of a suitable combination of orbital residuals of the 
laser-ranged LAGEOS
and LAGEOS II
satellites. While this observable is not affected by the orbital 
perturbation induced by  the zonal Earth solid and ocean tides, it is
sensitive to
those generated
by the tesseral
and sectorial tides. The assessment of their influence on the 
measurement of the parameter $\m_{LT}$, with which the 
gravitomagnetic effect is
accounted for, is the goal of this paper. After simulating
the combined residual curve by calculating accurately the mismodeling 
of the more effective tidal perturbations, it has been found that, 
while the
solid tides affect the recovery of $\m_{LT}$ at a level always  well 
below 1$\%$, for the ocean tides and the other long-period signals
$\D\m$ depends strongly on the observational period and the noise 
level: $\D\m_{tides} \simeq 2\%$  after 7 years. The aliasing
effect of $K_1\ l=3\ p=1$ tide and
SRP(4241) solar radiation pressure
harmonic, with periods longer than 4 years, on the perigee of LAGEOS 
II yield to a maximum systematic uncertainty on $\m_{LT}$ of
less than 4$\%$
over different observational
periods. The zonal 18.6-year tide does not affect the combined residuals.  }
\end{titlepage} \newpage
\setcounter{page}{1}
\vspace{0.2cm}
\baselineskip 14pt

\setcounter{footnote}{0}
\setlength{\baselineskip}{1.5\baselineskip}
\renewcommand{\theequation}{\mbox{$\arabic{equation}$}}

\section{Introduction}
According to Ciufolini (1996), the tiny general relativistic 
gravitomagnetic effect (Lense \& Thirring
1918) can be detected by analyzing the orbits
of the two laser-ranged LAGEOS and LAGEOS II satellites. The basic equation is
(Ciufolini 1996):
\eqi\dot y\equiv\d\dot\O_{exp}^{I}+k_1
\times\d\dot\O_{exp}^{II}+k_2\times\d\dot\og_{exp}^{II}=\m \times 
60.2.\lb{unolt}\eqf In it
$k_1=0.295$,
$k_2=-0.35$,
$\m$ is the
scaling
parameter, equal to 1 in general
relativity and 0 in classical mechanics,
to be determined and $\d\O_{exp}^{I},\ \d\O_{exp}^{II},\ \d\og_{exp}^{II}$ are
the orbital residuals, in milliarcseconds
(mas), calculated with the aid of orbit determination software like
UTOPIA [Center for Space Research, Univ. of Texas at Austin] or GEODYN
[NASA Goddard Space Flight Center], of the nodes of LAGEOS and
LAGEOS II and the perigee of
LAGEOS II. The residuals account for any unmodeled or mismodeled 
physical phenomena
acting on the observable
analyzed. By treating the gravitomagnetism
as an unmodeled physical effect, general relativity
forecasts its effect on \rfr{unolt}
to be:\eqi \dot y_{LT}\equiv
(31\ mas/y)+k_1\times (31.5\ mas/y)+k_2\times(-57\ mas/y)\simeq 60.2\
mas/y.\lb{duelt}\eqf

The determination of $\m$ is influenced by a great number of 
gravitational and nongravitational perturbations acting upon LAGEOS 
and
LAGEOS II.  Among the perturbations of gravitational origin a primary
role is played by Earth's solid and ocean
tides.

The combined residuals 
$y\equiv\d\O_{exp}^{I}+k_1\d\O_{exp}^{II}+k_2\d\og_{exp}^{II}$ allow 
to cancel out the static and dynamical contributions of
degree $l=2,4$ and order $m=0$ of the geopotential.  However, this is not
so for the tesseral ($m=1$) and sectorial ($m=2$) tides.

This paper aims to assess quantitatively, in a reliable and
plausible manner, how the solid and ocean Earth tides of order
$m=1,2$ affect the recovery of $\m$ by simulating the real residual 
curve and analyzing it. The $T_{obs}$ chosen
span ranges
from 4 years to 8 years: this
is so because 4 years is the
length of the latest time series actually analysed by Ciufolini et al.
(1998) and 8 years is the maximum length obtainable
today because LAGEOS II was launched in 1992.
The analysis
includes also the long-period
signals due to solar radiation pressure and the $l=3,\ m=0$ 
geopotential harmonic acting on the perigee of LAGEOS
II. In our case we have a signal built up with a secular trend, i.e. 
the Lense-Thirring effect, and a certain number of
long-period harmonics, i.e. the tesseral and sectorial tidal 
perturbations  and the other signals with known periodicities. The 
part of
interest for us is the
secular trend while the harmonic part represents the noise. We 
address the problem of how the harmonics affect the
recovery of the secular trend on given time spans $T_{obs}$ and various
samplings $\D t$.

Among the long-period signals we distinguish between those signals whose
periods are shorter than $T_{obs}$ and those with periods longer than 
$T_{obs}$. While the former average out if $T_{obs}$ is an integer 
multiple of their periods, the action of the latter is
more subtle since they could resemble  a trend over temporal 
intervals too short with respect to their periods. They must be
considered therefore as biases on the Lense-Thirring determination
affecting its recovery by means of their mismodeling. Thus, it is
of the utmost importance that we reliably assess their impact on the
determination of the trend from the gravitomagnetic effect. It would be
useful to direct the efforts of the community (geophysicists,
astronomers and space geodesists) towards the
improvement of our knowledge of those tidal constituents to which $\m$
turns out to be particularly sensitive (for the LAGEOS' orbits).

This investigation will quantify
unambiguously what one means with statements like: $\D\m_{tides}\leq
X\%\m_{LT}$. In this paper we shall try to put forward a
simple and meaningful approach. It must be pointed out though that it
is not a straightforward application of any exact or rigorously proven
method; on the contrary, it is, at a certain level, heuristic and
intuitive, but it has the merit of yielding reasonable and simple
answers and allowing for their critical discussion.

The paper is organized as follows. In
Section 2 the mismodeling of the various tidal orbit perturbations is 
worked out in order to use these
estimates in the simulation
procedure of the synthetic observable curve whose features are outlined in
Section 3. In Section 4 the effects of the harmonics with period
shorter than 5 years are examined by
comparing the least squares fitted values of $\m$ in two different 
scenarios: in the first one the
simulated curve is complete and the mathematical model we fit contains
all the most relevant signals  plus a straight line. In the second
one the harmonics are removed from the simulated curve which
is fitted by means of the  straight line only.  Section 5 address the 
topic of those harmonics with
periods longer than 5 years and Section 6 is devoted to the conclusions.
\section{The mismodeling of the tidal perturbations}
Since the observable is a residual
curve, if we want to simulate it we need reliable estimates of the
commission errors
of the various tides considered. They will take the place of real
residuals in the simulated data. In this section we address this
topic by calculating the mismodeled amplitudes of the solid and ocean 
tidal perturbations on the nodes of LAGEOS and LAGEOS II
and the perigee of LAGEOS II and comparing them to the corresponding 
gravitomagnetic perturbations over 4
years. A cutoff of $1\%$ has been set in order to obtain a 
preliminary evaluation of the importance of the various constituents.

Concerning the solid Earth tides, their perturbations on the node and 
the perigee of an artificial satellite, for a given
frequency $f$, are given by:
$$\D\O_f=\rp{g}{na^2\sqrt{1-e^2}\sin{i}}
\sum_{l=0}^{\infty}\sum_{m=0}^{l}
({\rp{R}{r}})^{l+1}\times$$\eqi\times A_{lm}
\sum_{p=0}^{l}
\sum_{q=-\infty}^{+\infty}\dert{F_{lmp}}{i}G_{lpq}\rp{1}{f_p}
k^{(0)}_{lm}(f)H_{l}^{m}\sin{\ga_{flmpq}},\lb{nodo}\eqf
$$\D\og_f=\rp{g}{na^2\sqrt{1-e^2}}
\sum_{l=0}^{\infty}\sum_{m=0}^{l} 
({\rp{R}{r}})^{l+1}\times$$\eqi\times A_{lm}\sum_{p=0}^{l}
\sum_{q=-\infty}^{+\infty}[
\rp{1-e^2}{e}F_{lmp}\dert{G_{lpq}}{e}-
\rp{\cos{i}}
{\sin{i}}\dert{F_{lmp}}{i}G_{lpq}]\rp{1}{f_p}
k^{(0)}_{lm}(f)H_{l}^{m}\sin{\ga_{flmpq}},\lb{perig}\eqf
while for the ocean tides we have:
$$ \D\O_f=\rp{1}{na^2\sqrt{1-e^2}\sin{i}}
\sum_{l=0}^{\infty}\sum_{m=0}^{l}\sum_{+}^{-}
({\rp{R}{r}})^{l+1}A_{lmf}^{\pm}\times $$
\eqi \times\sum_{p=0}^{l}
\sum_{q=-\infty}^{+\infty}\dert{F_{lmp}}{i}G_{lpq}\rp{1}{f_p}
\mc{\sin{\ga_{flmpq}^{\pm}}}{-\cos{\ga_{flmpq}^{\pm}}}^{l-m
\ even}_{l-m\  odd},\lb{nodoc}\eqf
$$ \D\og_f=\rp{1}{na^2\sqrt{1-e^2}}
\sum_{l=0}^{\infty}\sum_{m=0}^{l}\sum_{+}^{-}
({\rp{R}{r}})^{l+1}A_{lmf}^{\pm}\times $$
\eqi\times\sum_{p=0}^{l}\sum_{q=-\infty}^{+\infty}[
\rp{1-e^2}{e}F_{lmp}\dert{G_{lpq}}{e}-\rp{\cos{i}}
{\sin{i}}\dert{F_{lmp}}{i}G_{lpq}]\rp{1}{f_p} 
\mc{\sin{\ga_{flmpq}^{\pm}}}{-\cos{\ga_{flmpq}^{\pm}}}^{l-m
\ even}_{l-m\ odd}.\lb{pgoc}\eqf
The main source of uncertainties in such expressions
are the
$l=2$ Love numbers $k^{(0)}_{lm}(f)$,
the load Love numbers $k^{'}_{l}$, the coefficients $C^{+}_{lmf}$ and 
the orbital injection
errors $\d i$ affecting the inclination $i$.

As far as the solid tidal constituents,  by assuming  $\d i=0.5$ mas
(Ciufolini 1989), we have calculated $\vass{\derp{A(\O)}{i}}\d i$ and
$\vass{\derp{A(\og)}{i}}\d i$
for $055.565,\ K_1,\ $and $S_2$ which
are the most powerful tidal constituents in perturbing LAGEOS and 
LAGEOS II orbits.
The results
are of the order of $10^{-6}$ mas, so that we can neglect the
effect of
uncertainties in
the inclination determination.
Concerning the Love numbers $k^{(0)}_{2m}(f)$, we have assessed the 
uncertainties on them
by
calculating
for
certain tidal constituents the factor
$\d k_2/{k_2}$; ${k_2}$ is the average on the values released by the 
most reliable
models and $\d k_2$ is its standard deviation.
According to the recommendations of the Working Group of Theoretical 
Tidal Model of the
Earth Tide Commission 
(http://www.astro.oma.be/D1/EARTH$\_$TIDES/wgtide.html), in the 
diurnal band we have
chosen the values released by 
Mathews et al. (1995), McCarthy (1996) and the two sets by Dehant et 
al. (1999). For
the zonal and sectorial bands we have included also
the results of Wang (1994). The uncertainties
calculated in the Love numbers $k_2$ span from 0.5$\%$ to 1.5$\%$ for the tides
of
interest.  However, it must be noted
that the worst known Love numbers are those related to the zonal band 
of the tidal spectrum due to the
uncertainties on
the anelasticity of the
Earth' s mantle. These results were obtained in order to calculate
the mismodeled amplitudes of the solid
tidal perturbations $\d\O^{I},\
\d\O^{II}$ and $\d\og^{II}$; they were subsequently  compared with
the gravitomagnetic precessions over 4 years $\D\O^{I}_{LT}=124$ mas,
$\D\O^{II}_{LT}=126$ mas and $\D\og^{II}_{LT}=-228$ mas. In 
Tab.\ref{mismo1} we have quoted
only those tidal lines
whose mismodeled perturbative amplitudes are greater than $1\%$ of 
the gravitomagnetic perturbations. It
turns out
that only 055.565 18.6-year
and $K_1$ exceed this cutoff. Remember that the 18.6 year tide 
cancels out, so that we can conclude that the
uncertainties on the Love numbers affect the combined residuals at a 
level $\leq 1\%$.

Concerning the ocean loading,
the first calculations of the
load Love numbers $k^{'}_l$ are due to Farrell (1972). Pagiatakis 
(1990) in a first
step has recalculated $k^{'}_{l}$ for an elastic,
isotropic and
non-rotating Earth: for $l<800$ he
claims that his estimates differ from those by Farrell (1972), 
calculated with the same hypotheses,
by less
than $1\%$. Subsequently, he added to the
equations, one at a time, the effects of anisotropy,
rotation and dissipation; for low values of $l$, their effect on the
results of the calculations amount to
less than
1$\%$. For the ocean loading coefficients see also (Scherneck, 1991). It
has
been decided to calculate $\vass{\derp{A(\og)}{k^{'}_{l}}}\d k^{'}_{l}$
for the
perigee of LAGEOS II for
$K_1\ l=3\ p=1$ which turns
out to be the most powerful ocean tidal constituent acting on this orbital
element. First, we have calculated
mean and standard deviation of the values for $k^{'}_{3}$ by
Farrell
and Pagiatakis obtaining $\d k^{'}_{3}/k^{'}_{3}=0.9\%$, in
agreement with the estimates by Pagiatakis. Then, assuming
pessimistically
that the global effect of the departures
from these symmetric models results in
a total $\d k^{'}_{3}/k^{'}_{3}=2\%$, we have obtained 
$\d\og^{II}=5.5$ mas which corresponds to
$2\%$ of
$\D\og^{II}_{LT}$ over 4 years. Subsequently,
for this constituent and for all other tidal lines we have calculated 
the effect of the
mismodeling of $C^{+}_{lmf}$
as quoted in EGM96 (Lemoine et al. 1998).  In Tab.\ref{mismo2} we 
compare the so obtained mismodeled ocean
tidal
perturbations to those generated over 4 years
by the Lense-Thirring effect. It
turns out that the perigee of LAGEOS II is more sensitive to the 
mismodeling of the ocean part of the
Earth
response to the tide generating
potential. In particular, the effect of $K_1\ l=3\ p=1\ q=-1$ is 
relevant with a total $\d\og=
\vass{\derp{A(\og)}{C^{+}_{lmf}}}\d 
C^{+}_{lmf}+\vass{\derp{A(\og)}{k^{'}_{l}}}\d k^{'}_{l}$ of 64.5
mas
amounting to 28.3 $\%$ of $\D\og_{LT}$ over 4 years.
\begin{table}[ht!] \centering \tiny{\btab{|c|c|c|c|c|c|c|c|} \hline 
\multicolumn{8}{|c|}{ Mismodeled solid tidal
perturbations on nodes
of LAGEOS and LAGEOS II and perigee of LAGEOS II}\\
\hline
Tide & \multicolumn{7}{c|} { $\D\O^{I}_{LT}=124$ mas \ 
$\D\O^{II}_{LT}=126$ mas \ $\D\og^{II}_{LT}=-228$ mas \ $T_{obs}$=4 
years} \\
\cline{2-8}
       & $\d k_2/k_2$ ($\%$) & $\d\O^{I}$ (mas) & 
$\d\O^{I}/\D\O^{I}_{LT}$ ($\%$) & $\d\O^{II}$ (mas) &
$\d\O^{II}/\D\O^{II}_{LT}$ ($\%$) & $\d\og^{II}$ (mas) & 
$\d\og^{II}/\D\og^{II}_{LT}$ ($\%$)\\
\hline
055.565 & 1.5 & -16.5 & 13.3 & 30.3 & 24 & -21 & 9.2 \\
165.555 $K_1$& 0.5 & 9 & 7.2 & -2 & 1.6 & 10.2 & 4.4 \\
\hline
\etab}
\caption{\footnotesize{Mismodeled solid tidal perturbations on nodes 
$\O$ of LAGEOS  and LAGEOS II and the perigee $\og$ of
LAGEOS II compared to their gravitomagnetic precessions over 4
years. The percent variation refers  to the ratios of the mismodeled 
amplitudes of the tidal
harmonics to the gravitomagnetic perturbations over 4 years.}}
\label{mismo1}
\end{table}
\begin{table}[ht!] \centering \tiny{\btab{|c|c|c|c|c|c|c|c|} \hline 
\multicolumn{8}{|c|}{ Mismodeled ocean tidal
perturbations on nodes
of LAGEOS and LAGEOS II and perigee of LAGEOS II}\\
\hline
Tide & \multicolumn{7}{c|} { $\D\O^{I}_{LT}=124$ mas 
$\D\O^{II}_{LT}=126$ mas  $\D\og^{II}_{LT}=-228$ mas  $T_{obs}=4$ 
years} \\
\cline{2-8}
       & $\d C^{+}/C^{+}$ ($\%$)  & $\d\O^{I}$ (mas) & 
$\d\O^{I}/\D\O^{I}_{LT}$ ($\%$)  & $\d\O^{II}$ (mas) &
$\d\O^{II}/\D\O^{II}_{LT}$ ($\%$) & $\d\og^{II}$  (mas)& 
$\d\og^{II}/\D\og^{II}_{LT}$ ($\%$) \\
\hline
$S_{a}$ $l$=2 $p$=1 $q$=0& 6.7 & 1.37 & 1.1 & 2.5 & 1.9 & - & - \\
$S_{a}$ $l$=3 $p$=1 $q$=-1 & 10 & - & - & - & - & 11.4 & 5 \\
$S_{a}$ $l$=3 $p$=2 $q$=1 & 10 & - & - & - & - & 29.7 & 13 \\ \hline
$S_{sa}$ $l$=3 $p$=1 $q$=-1 & 27.2 & - & - & - & - & 6.2 & 2.7 \\
$S_{sa}$ $l$=3 $p$=2 $q$=1 & 27.2 & - & - & - & - & 9.8 & 4.3 \\ \hline
$K_1$ $l$=2 $p$=1 $q$=0& 3.8 & 5.9 & 4.7 & 1.3 & 1 & 6.75& 2.9\\
$K_1$ $l$=3 $p$=1 $q$=-1& 5.2 & - & - & - & - & 64.5 & 28.3\\
$K_1$ $l$=3 $p$=2 $q$=1& 5.2 & - & - & - & - & 18 & 7.9\\
$K_1$ $l$=4 $p$=2 $q$=0& 3.9 & - & - & 1.6 & 1.2 & - & -\\ \hline
$P_1$ $l$=3 $p$=1 $q$=-1& 18.5 & - & - & - & - & 5.3 & 2.3\\
$P_1$ $l$=3 $p$=2 $q$=1& 18.5 & - & - & - & - & 6.4 & 2.8\\ \hline
$K_2$ $l$=3 $p$=1 $q$=-1& 5.5 & - & - & - & - & 11.7 & 5\\
$K_2$ $l$=3 $p$=2 $q$=1& 5.5 & - & - & - & - & 4.8 & 2\\ \hline
$S_2$ $l$=3 $p$=1 $q$=-1& 7.1 & - & - & - & - & 6.9 & 3\\
$S_2$ $l$=3 $p$=2 $q$=1& 7.1 & - & - & - & - & 4.4 & 1.9\\ \hline
$T_2$ $l$=3 $p$=1 $q$=-1& 50 & - & - & - & - & 2.6 & 1.1\\ \hline
\etab}
\caption{\footnotesize{Mismodeled ocean tidal perturbations on nodes 
$\O$ of LAGEOS  and LAGEOS II and the perigee $\og$ of
LAGEOS II compared to their gravitomagnetic precessions over 4
years. The effect of the ocean loading has been neglected.  When the 
$1\%$ cutoff has
not been reached a - has been inserted. The values quoted for $K_1\ 
l=3\ p=1$ includes also the mismodeling in the ocean loading
coefficient
$k^{'}_{3}$ assumed equal to $2\%$. The percent variation refers  to 
the ratios of the mismodeled amplitudes of the tidal
harmonics to the gravitomagnetic perturbations over 4 years.}}
\label{mismo2}
\end{table}

\section{The simulated residual
signal}
The
first step of our strategy  was to generate with MATLAB a
time series which simulates, at the same length and
resolution,  the real residual
curve obtainable from
$y=\d\O_{exp}^{I}+k_1\d\O_{exp}^{II}+k_2\d\og_{exp}^{II}$ through, e.g.,
GEODYN.  This simulated curve (``Input Model" - IM in the following) was
constructed with:\\
$\bullet$ The secular Lense-Thirring trend as predicted by the 
general relativity\footnote{Remember that in the
dynamical
models of GEODYN II it was set purposely equal to 0 so that the
residuals absorbed (contained) entirely the relativistic effect
(Ciufolini et al. 1997).}\\
$\bullet$ A certain
number of sinusoids of the form $\d
A_k\cos{(\rp{2\p}{P_k}t+\f_k)}$ with known periods $P_k,\ k=1,..N$ 
simulating the mismodeled tides and other
long-period
signals which, to the level of assumed mismodeling,  affect the combined
residuals\\ $\bullet$ A  noise of
given amplitude which simulates the experimental errors on the 
laser-ranged measurements and, depending upon the
characteristics chosen for it, some other physical forces.\\
In a nutshell: \eqi
IM=LT+[mismodeled\
tides]+[other\ mismodeled\ long\ period\ signals]+[noise].\eqf   The 
harmonics included in the IM
are the following:\\
$\bullet\ K_1,\ l=2$ solid and ocean; node of LAGEOS (P=1043.67 days)\\
$\bullet\ K_1,\ l=2$ solid and ocean; node and perigee of LAGEOS II 
(P=-569.21 days)\\
$\bullet\ K_1,\ l=3,\ p=1$  ocean; perigee of LAGEOS II (P=-1851.9 days)\\
$\bullet\ K_1,\ l=3,\ p=2$ ocean; perigee of LAGEOS II (P=-336.28 days)\\
$\bullet\ K_2,\ l=3,\ p=1$ ocean; perigee of LAGEOS II (P=-435.3 days)\\
$\bullet\ K_2,\ l=3,\ p=2$  ocean; perigee of LAGEOS II (P=-211.4 days)\\
$\bullet\ 165.565,\ l=2$ solid; node of LAGEOS (P=904.77 days)\\
$\bullet\ 165.565,\ l=2$ solid: node and perigee of LAGEOS II 
(P=-621.22 days)\\
$\bullet\ S_2,\ l=2$ solid and ocean; node of LAGEOS (P=-280.93 days)\\    
$\bullet\ S_2,\ l=2$ solid and ocean; node and perigee of LAGEOS II 
(P=-111.24 days)\\
$\bullet\ S_2,\ l=3,\ p=1$ ocean; perigee of LAGEOS II (P=-128.6 days)\\    
$\bullet\ S_2,\ l=3,\ p=2$ ocean; perigee of LAGEOS II (P=-97.9 days)\\
$\bullet\ P_1,\ l=2$ solid and ocean; node of LAGEOS (P=-221.35 days)\\      
$\bullet\ P_1,\ l=2$ solid and ocean; node and perigee of LAGEOS II 
(P=-138.26 days)\\       
$\bullet\ P_1,\ l=3,\ p=1$ ocean; perigee of LAGEOS II (P=-166.2 days)\\ 
$\bullet\ P_1,\ l=3,\ p=2$ ocean; perigee of LAGEOS II (P=-118.35 days)\\
$\bullet\ Solar\ Radiation\ Pressure,$  perigee of LAGEOS II (P=-4241 days)\\
$\bullet\ Solar\ Radiation\ Pressure,$ perigee of LAGEOS II (P=657 days)\\  
$\bullet\ perigee\ odd\ zonal\ C_{30}$, perigee of LAGEOS II (P=821.79) \\
In the following the signals due to solar radiation pressure will be denoted as
SRP(P) where P is their period; the effects of the eclipses and Earth 
penumbra have not been accounted for.
Many of the periodicities listed above have been actually found in 
the Fourier spectrum of the real residual curve (Ciufolini et
al. 1997). Concerning $K_1\ l=3\ p=1$ and SRP(4241), see Section 5.

When the real data are collected they refer to a unique, unrepeatable 
situation characterized by  certain starting and ending dates
for $T_{obs}$. This means
that each analysis which could be carried out in the real world 
necessarily refers to a given set of initial phases
and noise for the residual curve
corresponding to the chosen observational
period; if the data are collected over the same $T_{obs}$ shifted 
backward or forward in time, in general,
such features of the curve will change. We neither know a priori when 
the next real experiment will be performed, nor which will be the
set of initial phases and the level of experimental errors accounted 
for by the noise. Moreover, maybe the dynamical models of the orbit
determination software
employed are
out of date with regard to the  perturbations acting upon satellites' 
orbits or do not include some of them at all. Consequently,
it would be
incorrect to work with a single simulated curve, fixed by an 
arbitrary set of $\d A_k,\ \f_k$ and noise, because it could refer to 
a
situation different from that in which, in the real world, the residuals
will actually correspond.

\begin{figure}[ht!] \begin{center}
\includegraphics*[width=17cm,height=12cm]{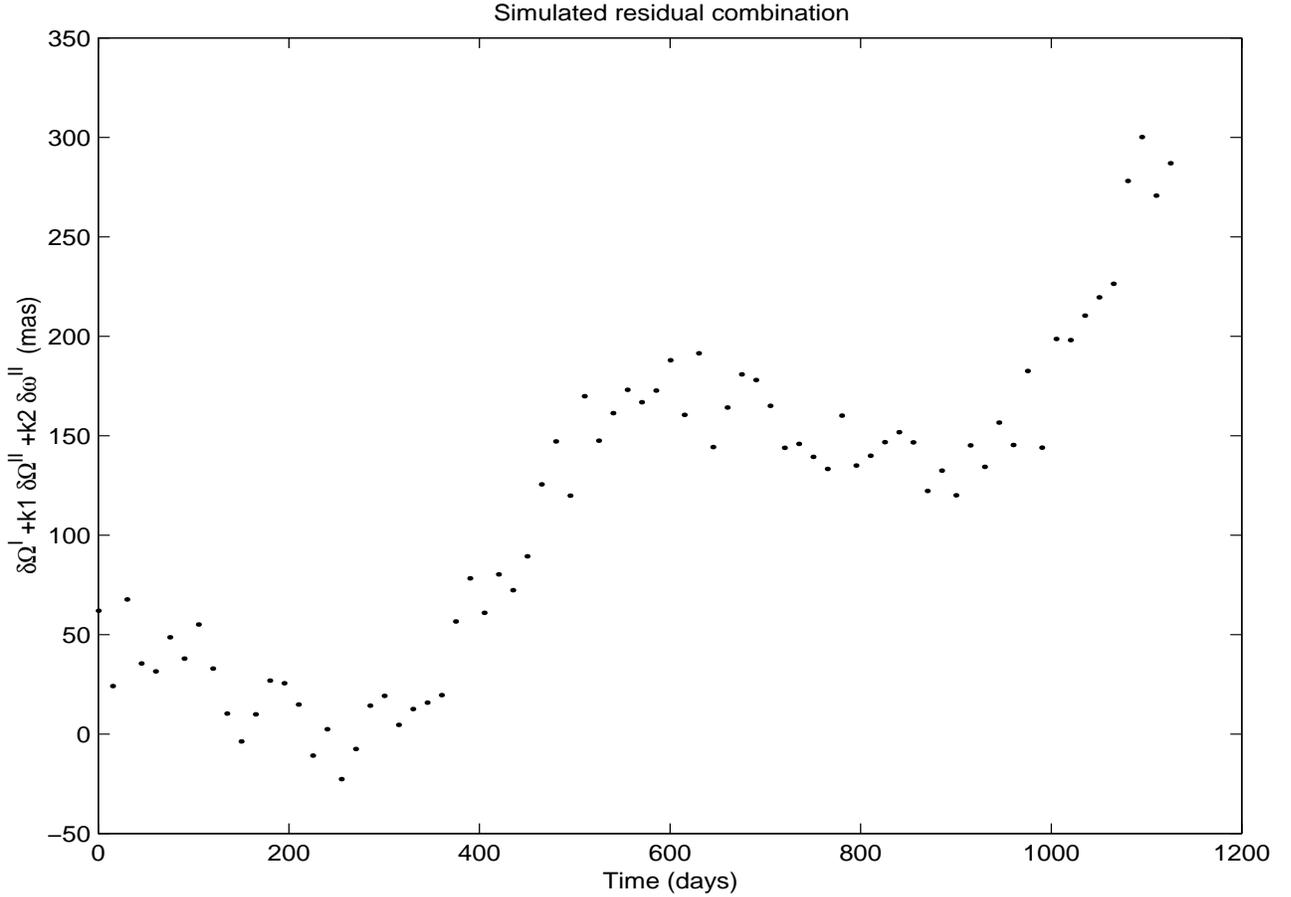}
\end{center}
\caption{\footnotesize Simulated residual curve. The time span is 3.1 
years, the
time step is 15 days and all the long period signals are included. The
random initial phases and the noise have been chosen in order to 
reproduce as closely as possible the residual curve of
(Ciufolini et al. 1997). The rms  of the simulated data amounts to 
almost 9 mas, the 5$\%$ of the
preticted gravitomagnetic value for
the combined residuals.}
\label{unof}
\end{figure}

  The need
for great flexibility in generating the IM becomes apparent: to
account for the entire spectrum of possibilities occurring when the real
analyses will be carried out. We therefore decided to build into the
MATLAB routine the capability to  vary  randomly the
initial phases $\f_k$, the noise and the amplitude errors $\d A_k$.
Concerning the error amplitudes
of the harmonics,
they can be  randomly varied so that $\d A_k \in [0,\ \d A_k^{nom}]$ 
where $\d A^{nom}_k$ is the nominal mismodeled amplitude
calculated
taking into account Tab.\ref{mismo1}
and Tab.\ref{mismo2}; it means that we assume  they are reliable 
estimates of the
differences between the real data and the dynamical models of the 
orbital determination softwares, i.e. of the residuals.
The MATLAB routine allows also the user to decide which
harmonic is to be included in the IM; it is also possible to choose 
the length of the
time series $T_{obs}$,
the sampling step $\D t$ and  the amplitude of the noise.
Fig.(\ref{unof}) shows a typical simulation over 3.1 years with $\D t$=
15 days and a given set of random initial phases and noise: all the 
long period signals have
been included with $\d A _k=\d A_k^{nom}$. It can be
compared to
the real residual curve released in (Ciufolini et al. 1997) for the same time
step and time span going from November 1992 to December 1995: 
qualitatively they agree very well.
We also calculated the root mean square for the IM simulated data, of 9
mas.

In order to assess
quantitatively this feature we proceeded as follows. First, over a 
time span of 3.1 years, we
fitted the IM
with a straight
line only, finding for a choice of random phases
and noise which qualitatively reproduces the curve shown in 
(Ciufolini et al. 1997), the value of 38.25 mas for the root
mean square of the post fit IM. The value quoted in (Ciufolini et
al. 1997) is 43 mas.  Secondly, as done in the cited paper, we fitted 
the complete IM with the LT plus a set of
long-period signals (see Section 4)  and then we subtracted the so 
adjusted harmonics from the original IM
obtaining a "residual"
simulated signal curve. The latter was subsequently fitted with a
straight line only, finding a rms post fit of 12.3 mas (it is nearly
independent of the random phases and the noise) versus 13 mas quoted in
(Ciufolini et al. 1997). 
A uniformly distributed  noise with nonzero average and amplitude of 50
mas was used (see Section 4).
These considerations suggest that the simulation procedure adopted is
reliable, replicates the real world satisfactorily and yields a good
starting point for conducting our sensitivity analyses.

\section{Sensitivity analysis}
A preliminary analysis was carried out in order to evaluate the
importance of the mismodeled solid tides on the recovery of $\m$.  We 
calculated the average:
\eqi \rp{1}{T_{obs}}\int_0^{T_{obs}}[solid\ tides]dt,\eqf where 
$[solid\ tides]$ denotes the analytical expressions of the mismodeled
solid tidal perturbations as given by \rfrs{nodo}{perig}. 
Subsequently, we compared it to the value of the gravitomagnetic
trend for the same $T_{obs}$.   The analyses were repeated by
varying $\D t$, $T_{obs}$ and the initial phases. They  have shown that
the mismodeled part of the solid Earth tidal spectrum is entirely
negligible with respect to the LT signal, falling always below $1\%$ of
the gravitomagnetic shift of the combined residuals accumulated over the
examined $T_{obs}$.    For the ocean tides and the other long-period
signals the problem was addressed in a different way. First, for a
given $\D t$ and different time series lengths, we included in the IM 
the effects of LT, the noise and the solid tides only: subsequently we
fitted it simply by means of a straight line. In a second step we have
simultaneously added, both to the IM and the fitting model (FM
hereafter), all the ocean tides, the solar radiation pressure SRP(657)
signal and the odd zonal geopotential harmonic. We then compared the
fitted values of $\m$ and $\d\m$ recovered from both cases in order to
evaluate $\D\m$ and $\D\d\m$. The least squares fits 
(Bard 1974; Draper $\&$ Smith 1981)
were performed by means of the  MATLAB routine ``nlleasqr" (see, e.g.,
http://www.ill.fr/tas/matlab/doc/mfit.html); as $\d\m$ we have 
assumed the square root of the diagonal covariance matrix element 
relative to the slope parameter.
  Note that the SRP(4241) has never been included in the FM (See 
Section 5), while $K_1\ l=3\
p=1$ has been included for $T_{obs}>5$ years.  
For both of the described scenarios  we have taken the average for 
$\m$ and $\d\m$ over
1500 runs performed by varying
randomly the initial phases, the noise and the amplitudes of the 
mismodeled signals in order to account for all possible
situations occurring in the real world, as pointed out in the 
previous section. The large number of repetitions
was chosen in order to avoid that statistical fluctuations in the 
results could ``leak" into
$\D\ol{\m}$ and $\D\ol{\d\m}$ and
corrupt them at the level of $1\%$. With 1500 runs the standard
deviations on $\ol{\m}$ and $\ol{\d\m}$ are of the
order of $10^{-3}$ or less, so that we can reliably use the results 
of such averages for our
comparisons of $\ol{\m}$(no tides) vs $\ol{\m}$(all tides).

Before implementing such strategy for different $T_{obs},\ \D t$ and 
noise we carefully analyzed the $\D t=15$ days, $T_{obs}$=4
years experiment considered in (Ciufolini et
al. 1998) by trying to obtain the quantitative features outlined there,
so that we start from a firm and reliable basis. This goal was
achieved  by proceeding as outlined in the previous Section and 
adopting a uniformly distributed random noise with an amplitude of 35 
mas. 

\begin{figure}[ht!]
\begin{center}
\includegraphics*[width=17cm,height=12cm]{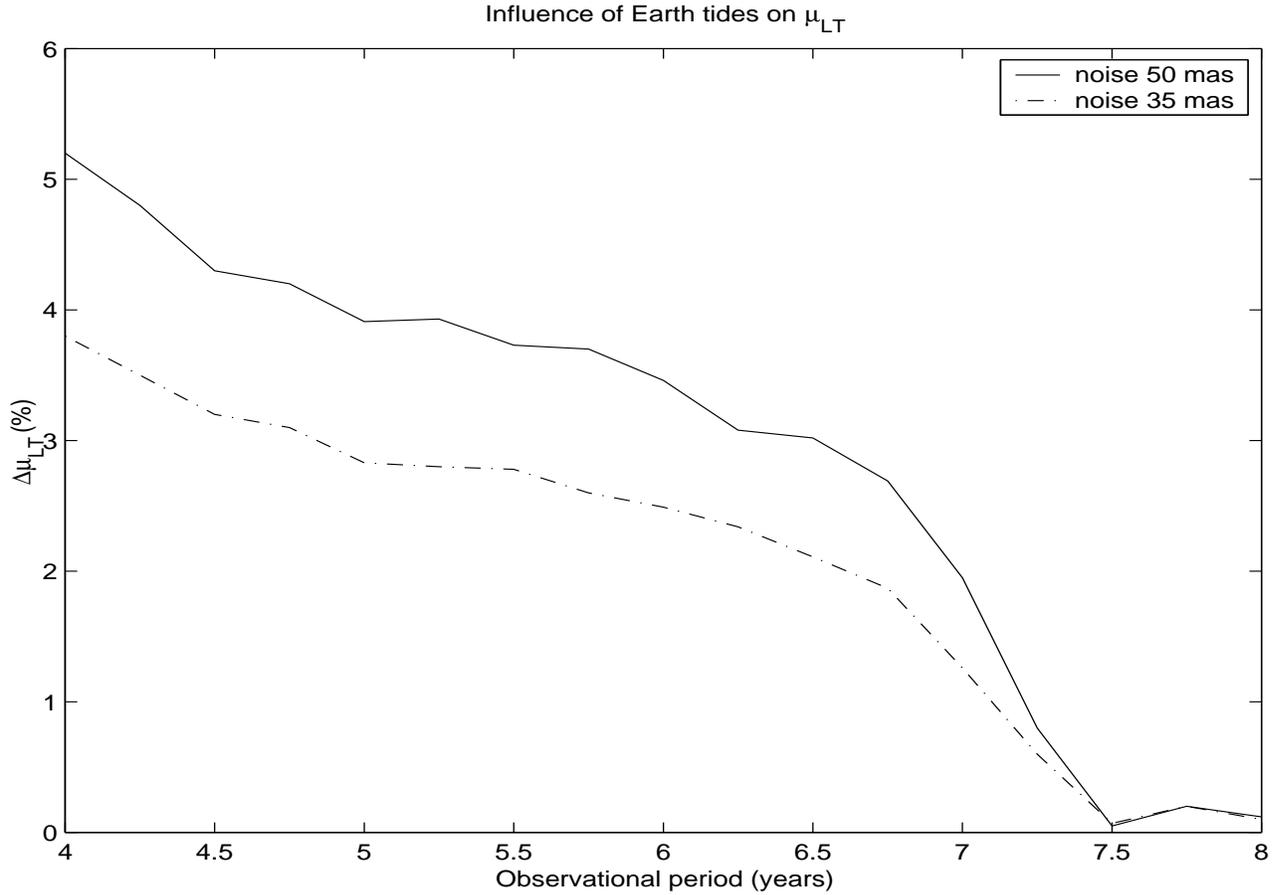}
\end{center}
\caption{\footnotesize
Effects of the long period signals on the recovery of $\m_{LT}$ for 
$\D t$=15 days and different choice of uniform random
noise. Each point in the curves represents an average over 1500 runs 
performed by varying randomly the initial phases and the
noise' s pattern. $\D\m$ is the
difference
between the least squares fitted value of $\m$ when both the IM and 
the FM includes the LT plus all the harmonics and that obtained
without any harmonic both in the IM and in the FM.} \label{dmu}
\end{figure}

Fig.\ref{dmu} shows the results for $\D\ol{\m}$ obtained with $\D 
t=15$ days and a
uniform random noise with amplitudes of 50 mas and 35 mas
corresponding to the
characteristics of the real curves in (Ciufolini et al. 1997; 1998). 
Note that our estimates for the case $T_{obs}=4$ years
almost coincide with those by Ciufolini et al. (1998) who claim 
$\D\m_{tides}\leq 4\%$.
Note that for $T_{obs}>7$ years the effect of tidal perturbations errors
falls around $2\%$. By choosing different $\D t$ does not introduce
appreciable modifications to the results presented here. This fact was
tested by repeating the set of runs with $\D t=7$ and $22$ days.

Up to this point we dealt with the entire set of long-period signals 
affecting the combined residuals; now we
ask if it is possible  to assess individually the influence of
each tide on the recovery of LT. We shall focus our attention on
the case $\D t=15$ days, $T_{obs}$=4 years.

In order to evaluate the influence of each tidal constituent on the 
recovery of $\m$ two complementary approaches could be
followed in principle. The first one consists in starting without any 
long-period signal either in the IM or in the FM,
and subsequently adding to them one harmonic at a time, while neglecting
all the others. In doing so it is implicitly assumed that each 
constituent is mutually uncorrelated with any other one present in 
the signal. In fact, the matter is
quite different since if for the complete model case we consider
\begin{figure}[ht!]
\begin{center}
\includegraphics*[width=17cm,height=12cm]{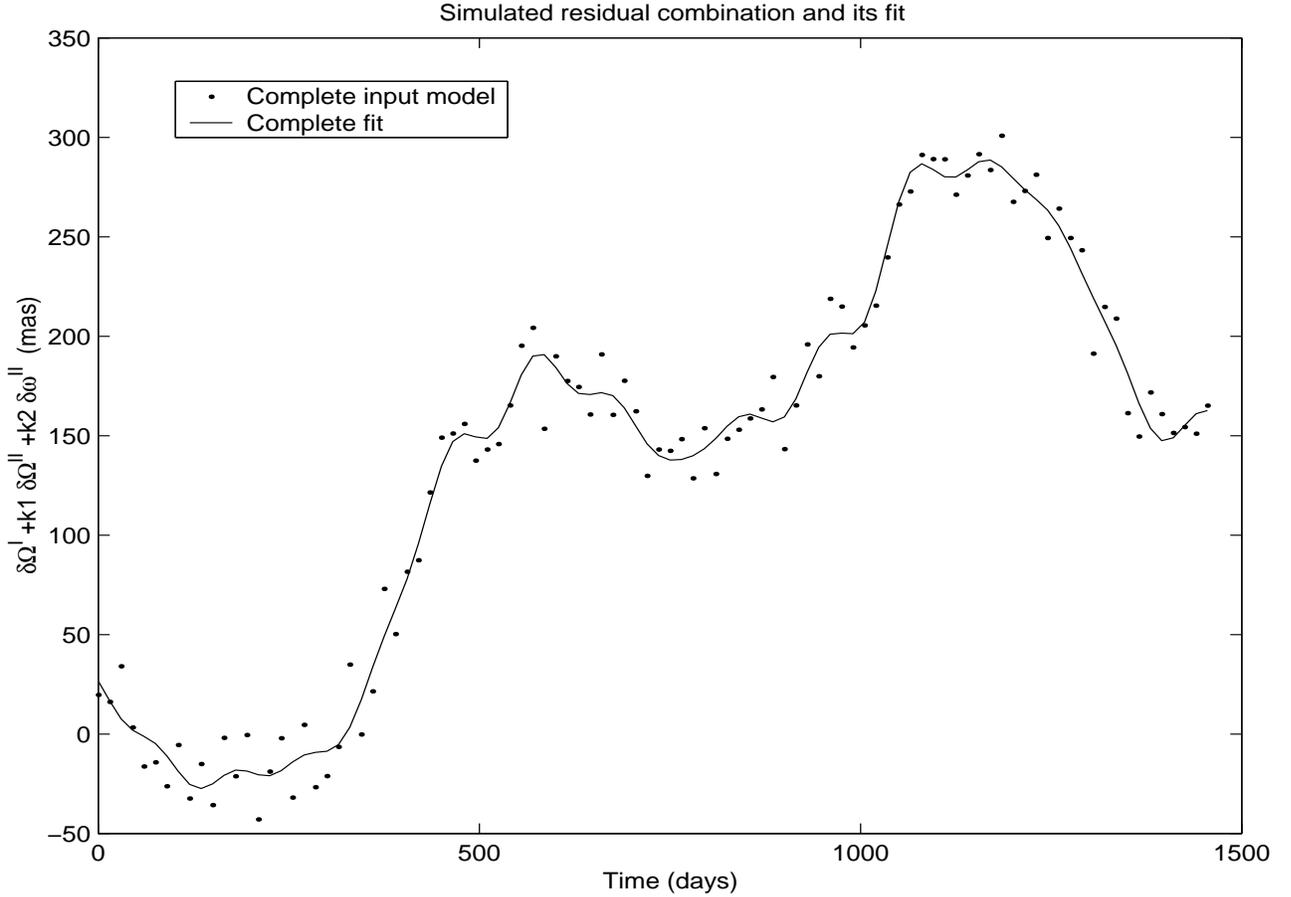}
\end{center}
\caption{\footnotesize{Simulated residual curve and related fit. The 
time span is 4 years, the
time step is 15 days, the simulated data RMS amounts to 9 mas and all the long
period signals are included in the simulated data. The
initial phases and the noise are random.}} \label{duef}
\end{figure}
the covariance and correlation matrices of the FM adjusted parameters it
turns out that the LT is strongly correlated, at a level of
$\vass{Corr(i,j)}> 0.9$, with certain harmonics, which happen to be
mutually correlated too. These are:\\
$\bullet\ K_1,\ l=2$ (P=1043.67 days; 1.39 cycles completed)\\
$\bullet\ K_1,\ l=2$ (P=-569.21 days; 2.56 cycles completed)\\
$\bullet\ K_1,\ l=3,\ p=2$ (P=-336.28 days; 4.34 cycles completed)\\
$\bullet\ K_2,\ l=3,\ p=1$ (P=-435.3 days; 3.35 cycles completed)\\
$\bullet\ Solar\ Radiation\ Pressure,$ (P=657 days; 2.22 cycles completed)\\  
$\bullet\ perigee\ odd\ zonal\ C_{30}$, (P=821.79; 1.77 cycles completed)\\
Their FM parameters are indeterminate in the sense that the
values estimated are smaller than the relative uncertainties assumed to be
$\sqrt{Cov(i,i)}$. On the contrary, there are other signals which are 
poorly correlated to the LT and whose reciprocal
correlation too is very low and that are well determined.
These are:\\
$\bullet\ K_2,\ l=3,\ p=2$ (P=-211.4 days;  6.91 cycles completed)\\
$\bullet\ S_2,\ l=3,\ p=1$ (P=-128.6 days;  11.3 cycles completed)\\    
$\bullet\ S_2,\ l=3,\ p=2$ (P=-97.9 days;  14.9 cycles completed)\\  
$\bullet\ P_1,\ l=3,\ p=1$ (P=-166.2 days; 8.78 cycles completed)\\ 
$\bullet\ P_1,\ l=3,\ p=2$ (P=-118.35 days; 12.3 cycles completed)\\
In Fig.(\ref{duef}) the complete IM and
FM are shown for a given choice of the initial phases, uniform noise 
with amplitude of 50 mas,  $T_{obs}=4$ years and $\D t=$
days.
It is interesting
to note that over $T_{obs}=4$ years the signals uncorrelated with the 
LT have in general described many
complete
cycles, contrary to these correlated with LT. This means that  the 
signals that average out over $T_{obs}$
decorrelate with LT, allowing
the
gravitomagnetic
trend to
emerge clearly against the background, almost not affecting the LT
recovery. 
This feature has been tested as follows.
In a first step, all the uncorrelated tides have been removed from 
both the IM (50 mas uniform random noise)
and
the FM, leaving only the correlated tides in. The runs were then 
repeated,  all other things being equal,  the
following values were recorded: $\ol{\m}=1.2104,\
\ol{\d\m}=0.1006$ with a variation with respect to the complete model 
case ($\ol{\m}=1.2073$, $\ol{\d\m}=0.1295$ ) of:
$\D\ol{\m}\simeq 0.3\%,\ \D\ol{\d\m}\simeq 2.9\%$. Conversely, if all 
the signals
with strong
correlation  are removed from the simulated data and
from the FM, leaving only the uncorrelated ones in, we obtain: 
$\ol{\m}=1.1587,\
\ol{\d\m}= 0.0186$ with  $\D\ol{\m}\simeq 4.8\%,\ \D\ol{\d\m}\simeq 
11\%$. Note that the sum of both contributions for $\D\ol{\m}$
yields exactly the overall value $\D\ol{\m}=5.2\%$ as obtained in the 
previous analysis (cfr. Fig.\ref{dmu} for $T_{obs}=4$ years).
This highlights the importance of certain long period signals in 
affecting the recovery
of LT and justify the choice of treating them simultaneously as it 
was done in obtaining Fig.\ref{dmu}. Moreover, it clearly indicates
that the
efforts of the scientific community should be focused on the 
improvement of the knowledge of these tidal constituents. 
It has been tested that for $T_{obs}=8$ years all such ``geometrical" 
correlations among the LT and the harmonics nearly
disappear, as it would be intuitively expected.
\section{The effect of the very long periodic harmonics}
In this Section we shall deal with those signals whose periods are 
longer than 4 years which could  corrupt the recovery of LT
resembling superimposed trends if their periods are considerably 
longer than the  adopted time series length.

We have shown the existence of a very long periodic ocean tidal 
perturbation acting upon the perigee of LAGEOS II. It is
the $K_1\ l=3\ p=1\ q=-1$
constituent with period $P=1851.9$ days (5.07 years)  and nominal 
amplitudes of $-1136$ mas (Iorio 2000). In (Lucchesi 1998), for the
effect of the direct solar
radiation pressure on the perigee of LAGEOS II, it has been 
explicitly calculated, by neglecting the effects of the eclipses, a
signal SRP(4241) with
$P=4241$ days (11.6 years) and nominal
amplitude of
6400 mas.
The mismodeling on these harmonics, both of the form $\d 
A\sin{(\rp{2\p}{P}t+\f)}$, amounts to 64.5 mas for the tidal
constituent, as shown in Section 1, and to 32 mas for SRP(4241), 
according to (Lucchesi 1998).

About the
actual presence of such semisecular harmonics in  the spectrum of the 
real combined residuals, it must be pointed out that, over
$T_{obs}=3.1$ years
(Ciufolini et al. 1997), it is not possible that so low frequencies 
could be resolved by Fourier analysis.
Indeed, according to (Godin 1972; Priestley 1981), when a signal is 
sampled over a finite time interval
$T_{obs}$ it induces a sampling also in the
spectrum. The lowest
frequency that can be resolved is:
\eqi f_{min}=\rp{1}{2T_{obs}},\lb{freqz}\eqf i.e. a harmonic must 
describe, at least, half a cycle over
$T_{obs}$ in order to be detected in the spectrum. $f_{min}$ is 
called elementary frequency band and it also represents
the minimum separation
between two frequencies in order to be resolved. For our two signals 
we have $f(K_1)=5.39\cdot 10^{-4}$ cycles per day (cpd)
and $f(SRP)=2.35\cdot 10^{-4}$
cpd; over 3.1 years $f_{min}=4.41\cdot 10^{-4}$ cpd. This means that 
the two harmonics neither can be resolved as
distinct nor they can be found
in the spectrum at all. In order to resolve them we should wait for 
$T_{obs}=4.5$ years which corresponds to their separation
$\D f=3.04\cdot 10^{-4}$ cpd, according to \rfr{freqz}.
In view of the potentially large aliasing effect of these two harmonics on
the LT it was
decided  to include both $K_1\ l=3\ p=1$ and SRP(4241) in the simulated
residual curve at the level of mismodeling
claimed before.

\begin{figure}[ht!]
\begin{center}
\includegraphics*[width=17cm,height=12cm]{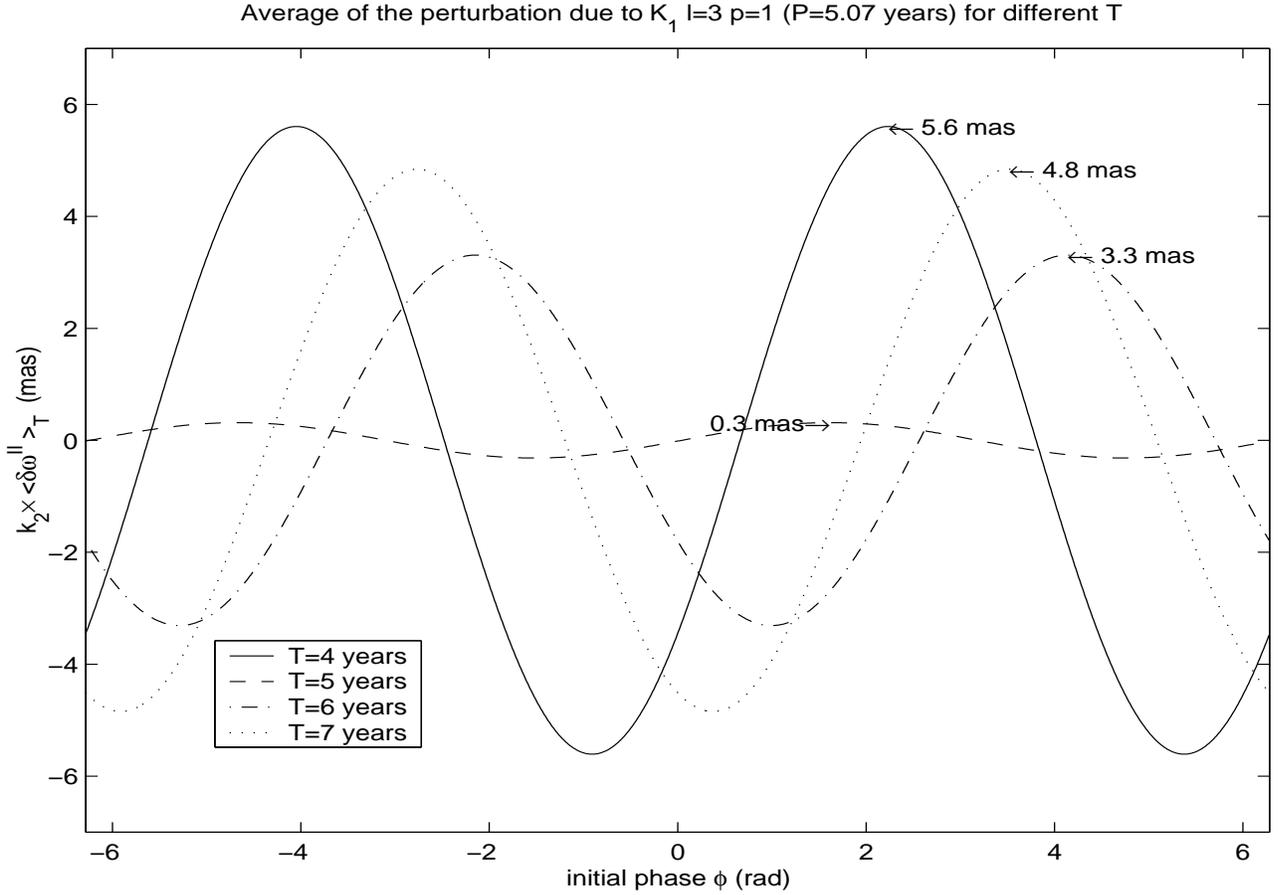}
\end{center}
\caption{\footnotesize Average $k_2<\d\og^{II}>_T$ over different 
$T_{obs}$ of the perturbation induced on the combined residuals by
the mismodeled harmonic $K_1\ l=3\ p=1$. In
general, it depends on
the initial phase $\f$. It has been calculated a mismodeling of 64.5 
mas from a nominal amplitude of 1136 mas. It is mainly due to the
$C^+_{31f}$ ocean tidal coefficient and the load Love number $k^{'}_3$.}
\label{avk1}
\end{figure}

\begin{figure}[ht!]
\begin{center}
\includegraphics*[width=17cm,height=12cm]{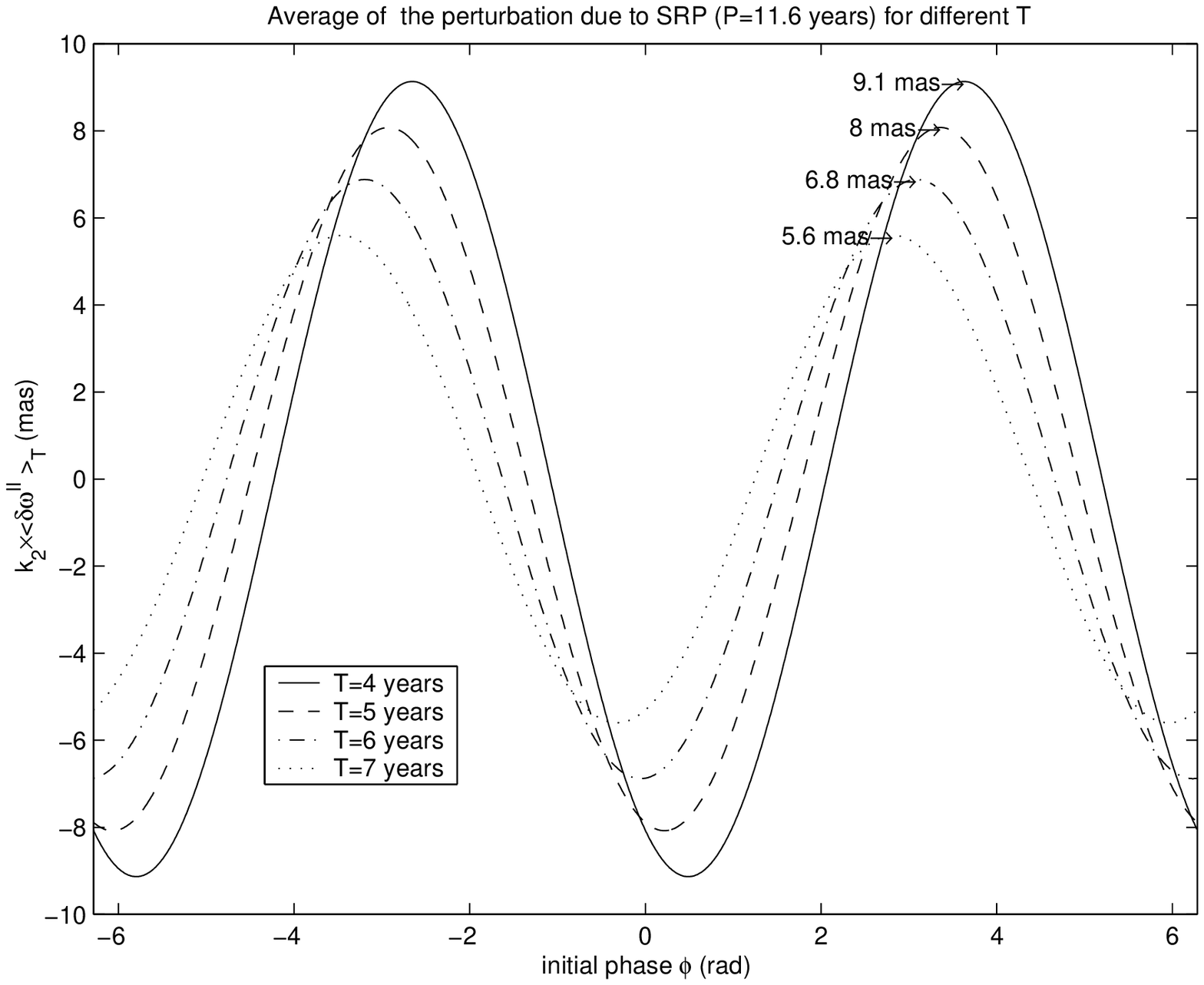}
\end{center}
\caption{\footnotesize Average $k_2<\d\og^{II}>_T$ over different 
$T_{obs}$ of the perturbation induced on the combined residuals by
the mismodeled harmonic
SRP(4241). In
general, it depends on
the initial phase $\f$. It has been assumed a $0.5\%$ level of 
mismodeling mainly due to the reflectivity coefficient $C_R$ of
LAGEOS II. The effects of the eclipses have been neglected.}
\label{avsrp}
\end{figure}

In order to obtain an upper bound of their contribution to the 
systematic uncertainty on  $\m_{LT}$ we proceeded as
follows. First, we calculated the temporal average
of the perturbations induced on the combined residuals by the two
harmonics over different $T_{obs}$ according to:
\eqi
I=\rp{1}{T_{obs}}\int_{0}^{T_{obs}}k_2\d
A \sin{(\rp{2\p}{P} t+\f)}dt
=k_2\rp{\d 
A}{\t}(\cos{\f}+\sin{\f}\sin{\t}-\cos{\f}\cos{\t}),\lb{ave}\eqf with 
$\t=2\pi\rp{T_{obs}}{P}$. The results
are
shown in Fig.\ref{avk1} and Fig.\ref{avsrp}.

  It can be seen that, for given $T_{obs}$,  the averages are periodic 
functions of the initial phases $\f$ with period $2\p$.
Since the gravitomagnetic trend is  positive and  growing in time, we 
shall consider only the positive values of the temporal
averages
corresponding to those portions of sinusoid which are themselves 
positive. Moreover, note that, in this case, one
should consider if the perturbation' s arc
is
rising or falling over
the considered $T_{obs}$: indeed, even though the corresponding 
averages could be equal, it is only in the first case
that the sinusoid has an aliasing effect on the
LT trend. The values
of $\f$ which maximize the averages were found numerically and for
such  values  the maxima of the averages were calculated.
Subsequently, these results in mas were compared to the amount of
the predicted gravitomagnetic shift accumulated over the chosen $T_{obs}$
by the combined
residuals $y_{LT}=60.2$ (mas/y)$\times T_{obs}$ (y). The results are 
shown in Tab.\ref{mediak1} and Tab.\ref{mediasrp} and, as previously
outlined, represent a
pessimistic estimate.
\begin{table}[ht!] \centering
\btab{||c||c|c|c||} \hline
\multicolumn{4}{||c||}{Averaged mismodeled $K_1\ l=3\ p=1$}\\
\hline                                   
$T_{obs}$ (years) & \multicolumn{3}{c||} {$P(K_1\ l=3\ p=1)=5.07$ 
years} \\ \cline{2-4}
       & max $k_2<\d\og^{II}>$ (mas) & $y_{LT}$ (mas) & max 
$k_2<\d\og^{II}>/y_{LT}\  (\%)$ \\
\hline
4 & 5.6 & 240.8 & 2.3  \\ \hline
5 & 0.3 & 301 &  0.09\\ \hline
6  & 3.3 & 361.2 & 0.9 \\ \hline
7 & 4.8 &  421.4 & 1.1\\ \hline        
\etab
\caption{\footnotesize{Effect of the averaged mismodeled harmonic 
$K_1\ l=3\ p=1$ on the Lense-Thirring trend for different
$T_{obs}$. In order to obtain upper limits the maximum value for the 
average of the tidal constituent has been taken,
while for the gravitomagnetic effect it has been simply taken the 
value $\dot y_{LT}\times T_{obs}$.}}\label{mediak1}
\end{table}

\begin{table}[ht!] \centering
\btab{||c||c|c|c||} \hline 
\multicolumn{4}{||c||}{Averaged mismodeled $SRP(4241)$}\\
\hline
$T_{obs}$ (years) & \multicolumn{3}{c||} {$P(SRP)=11.6$ years} \\ 
\cline{2-4}                  
       & max $k_2<\d\og^{II}>$ (mas) & $y_{LT}$ (mas) & max 
$k_2<\d\og^{II}>/y_{LT}\  (\%)$ \\
\hline               
4 & 9.1 & 240.8 & 3.7  \\ \hline                            
5 & 8 & 301 &  2.6\\ \hline                                      
6  & 6.8 & 361.2 & 1.8 \\ \hline                                    
7 & 5.6 &  421.4 & 1.3\\ \hline                                 
\etab
\caption{\footnotesize{Effect of the averaged mismodeled harmonic 
$SRP(4241)$ on the Lense-Thirring trend for different
$T_{obs}$. In order to obtain upper limits the maximum value for the 
average of the radiative harmonic has been taken,
while for the gravitomagnetic effect it has been simply taken the 
value $\dot y_{LT}\times T_{obs}$.}}\label{mediasrp}
\end{table}
It is interesting to
note that an analysis over $T_{obs}=5$ years, that should not be much 
more demanding than the already performed works,
could cancel out the
effect of the $K_1\ l=3\ p=1$ tide; in this scenario the effect of 
SRP(4241) should amount, at most to $2.6\%$ of the LT
effect.
Note also that for $T_{obs}=7$ years, a time span sufficient for the 
LT to emerge on the background of  most of the other tidal
perturbations, the estimates of Fig.\ref{dmu}
are compatible  with
those of Tab.\ref{mediak1} and Tab.\ref{mediasrp} which predict an
upper contribution of $1.1\%$ from the $K_1 \ l=3\ p=1$ and $1.3\%$
from SRP(4241) on the LT parameter $\m$.

An approach, similar
to that used in (Vespe 1999) in order to assess the influence of the eclipses and Earth penumbra 
effects on the perigee of
LAGEOS II has been
applied also to our case. It consists in fitting with a straight line only the 
mismodeled perturbation to be considered and , subsequently,
comparing
the slope of such
fits to that due to gravitomagnetism which is equal to 1 in units of 
60.2 mas/y. We have applied this method to $K_1 \ l=3\ p=1$
and
SRP(4241) with the already cited mismodeled amplitudes and by varying 
randomly the initial phases $\f$ within $[-2\p,\ 2\p]$. The
mean value of
the fit' s slope, in units of 60.2 mas/y, over 1500 runs is very 
close to zero. This agrees with Fig.\ref{avk1} and Fig.\ref{avsrp}
which tell
us that the temporal
averages of $K_1\ l=3\ p=1$ and SRP(4241) are periodic functions of 
$\f$ with period of $2\p$ and, consequently, have
zero mean value. Concerning the upper limits of $\D\m$ derived  from 
these runs, they agree with those released in Tab.\ref{mediak1} and
Tab.\ref{mediasrp} up to  1-2 $\%$.

\begin{figure}[ht!]
\begin{center}
\includegraphics*[width=17cm,height=12cm]{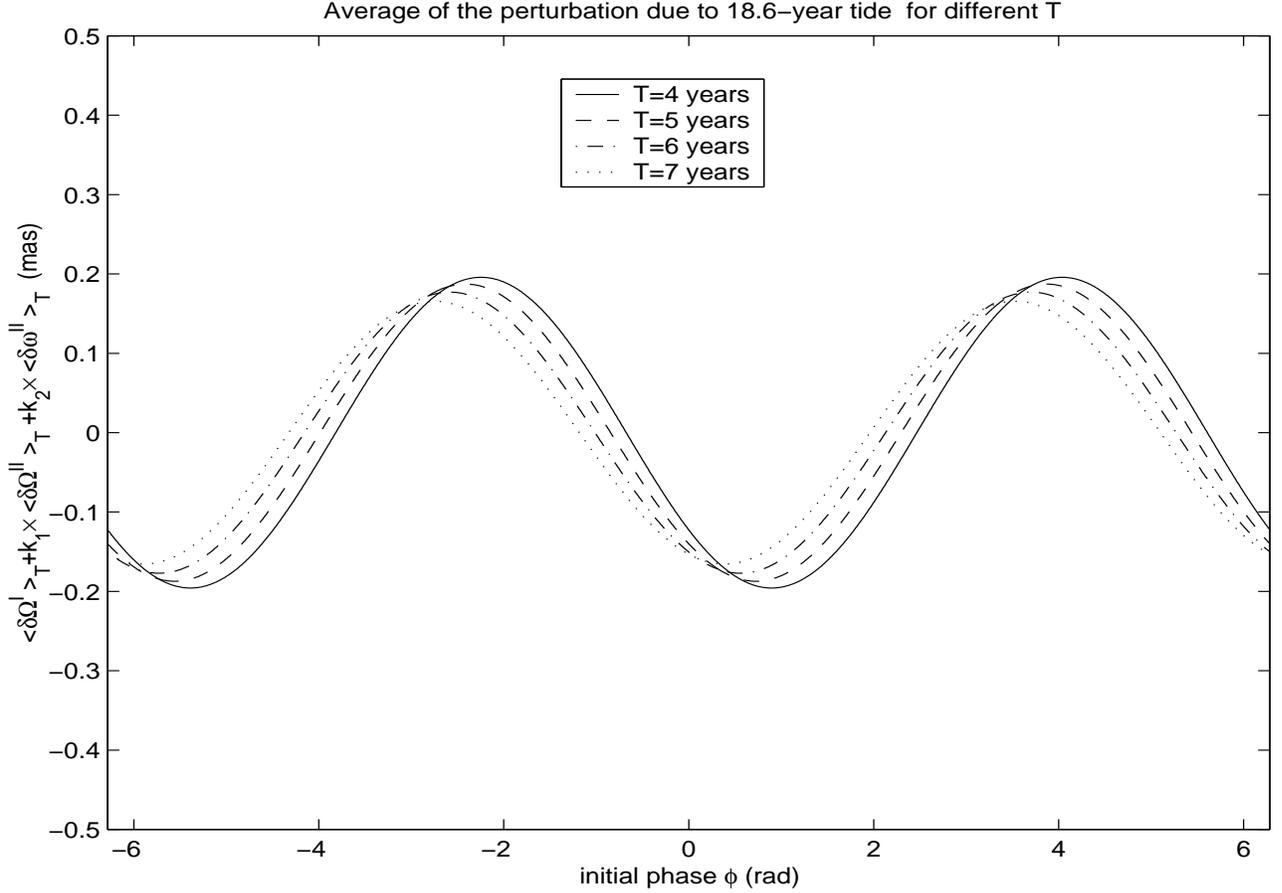}
\end{center}
\caption{\footnotesize Average 
$<\d\O^{I}>_T+k_1<\d\O^{II}>_T+k_2<\d\og^{II}>_T$ over different 
$T_{obs}$ of the perturbation induced on
the combined residuals by
the mismodeled 18.6-year tide. In
general, it depends on
the initial phase $\f$. It has been assumed a $1.5\%$ level of 
mismodeling on the Love number $k_2$ mainly due to the anelasticity 
of the
Earth' s mantle behaviour.}
\label{lungalunga}
\end{figure}

The method of temporal averages can be successfully applied also for 
the $l=2\ m=0$ 18.6-year tide. It is a very long period zonal tidal
perturbation which could potentially reveal itself as the most 
dangerous in aliasing the results for $\m$ since its nominal 
amplitude is
very
large and
its period is much longer than the $T_{obs}$ which could be adopted for
real analyses. Ciufolini (1996) claims that the combined
residuals have the merit
of cancelling out
all the static and dynamical geopotential' s contributions of degree 
$l=2,4$ and order $m=0$, so that the 18.6-year tide would
not create problems.
This topic was quantitatively addressed in a preliminary
way in (Iorio 2000). Indeed, in order to make
comparisons
with
other works, we have simply calculated for $T_{obs}=1$ year the
combined residuals with only the
mismodeled amplitudes of the
18.6-year tidal perturbations on the nodes of LAGEOS and LAGEOS II 
and the perigee of LAGEOS II. This means that
the dynamical pattern over the time span of such important tidal
perturbation was not investigated. We did this by
calculating the average over different $T_{obs}$. The
results are in Fig.\ref{lungalunga}  which shows clearly that the the
18.6-year tide does not affect at all the estimation of $\m$  if we adopt
as observable the combined residuals proposed by Ciufolini. Indeed, for
$T_{obs}=4$ years, the average effect will reach, at most, $0.08\%$ of the
gravitomagnetic shift over the same time span.

This feature of the 18.6-year tide is confirmed also by fitting with a
straight line only the sinusoid of this perturbation on the
combined residual: the adjusted slope amounts, at most,  to less than
1$\%$ of the gravitomagnetic effect for different $T_{obs}$.
\section{Conclusions}
In this paper we have evaluated quantitatively
the effects of mismodeling the orbital
perturbations due to the tesseral and sectorial ocean
and solid Earth tides on the
combined nodes and perigee residuals of LAGEOS and LAGEOS
II as proposed in (Ciufolini 1996) in order to detect the Lense-Thirring
effect.

On one hand, by simulating
the real residual curve in order
to reproduce as closely as possible
the results obtained for the $\D t=15$ days $
T_{obs}=4$
years
scenario published in (Ciufolini
et al. 1998) it has been
possible to refine and detail the estimates for it. On the other hand,  this
procedure also extends them
to longer observational periods in view of new, more sophisticated
analyses to be completed in the near
future based on real data analyzed with the orbit determination software
GEODYN II in collaboration with the teams from the Joint Center for 
Earth Systems Technology at NASA Goddard Space
Flight Center and
at the University of Rome
La Sapienza. Since such numerical analyses are very demanding in terms
of both time
employed and 
results analysis burden, it should be
very useful to have  a priori
estimates which could better direct the work. This could be done,
e.g., by identifying
which tidal constituents
$\m_{LT}$ is  more or less sensitive
to in order to seek improved dynamical models for use in GEODYN. 

As far as the perturbations
generated by the solid Earth tides, the high level of accuracy with 
which they are known has yielded a contribution
to the systematic errors on $\m_{LT}$ which falls well below 1$\%$, 
so that they are of no concern
at present.

Concerning the ocean tidal
perturbations and the other long-period
harmonics, for those  whose periods are shorter than 4 years, the 
role played by $T_{obs}$, $\D t$ and the noise has been
investigated. It
turned out that $\D t$ has no discernible effect on the adjusted value of
$\m$, while $T_{obs}$ is very important and so is the noise.
The main results for these are summarized in Fig.\ref{dmu}, which 
tells us that the entire set of long-period signals, if properly
accounted for in building up
the residuals, affect the recovery
of the Lense-Thirring effect at a level not worse than $4\%-5\%$ for
$T_{obs}=4 $ years; the error contribution diminishes to about $2\%$
after 7 years of observations.

We have also shown
which tides are strongly anticorrelated and correlated with the 
gravitomagnetic trend over 4 years
of observations. The experimental
and theoretical efforts should concentrate on improving these
constituents in particular.
This geometrical correlation tends to
diminish as $T_{obs}$ grows. This can be
intuitively
recognized by noting that the
longer  $T_{obs}$ is, the larger the number of cycles these periodic 
signals are sampled over and cleaner the way in which the
secular Lense-Thirring trend emerges against
the background ``noise".

The ocean tide constituents $K_1\ l=3\ p=1$ and
the solar radiation pressure harmonic SRP(4241) generate 
perturbations on the perigee of LAGEOS II 
with periods of 5.07 years and 11.6 years
respectively, so that they act on the Lense-Thirring effect as biases 
and corrupt its determination with
the related mismodeling: indeed they
may resemble  trends if $T_{obs}$ is shorter than their periods. They
were included
in the simulated residual curve and their effect was evaluated in
different ways with respect to the other tides. An upper bound was
calculated for their action and it turns out that they contribute to
the systematic uncertainty on the recovery of $\m_{LT}$ at a level of 
less than 4$\%$  depending on
$T_{obs}$ and the initial phases $\f$. The results
are summarized in Tab.\ref{mediak1} and Tab.\ref{mediasrp}. An 
observational period of 5 years,
which seems to be a reasonable choice in terms of time scale and
computational burden, allows to average out the effect of the $K_1\ l=3\ p=1$.

The strategy followed for the latter harmonics has been extended also 
to the $l=2\ m=0$ zonal
18.6-year tide. Fig.\ref{lungalunga} confirms the claim in (Ciufolini 
1996) that it does not affect the combined residuals.

In conclusion, the strategy presented here could be used as follows. 
Starting from a simulated residual curve based on the state of
art of the real
analyses
performed until now, it provides helpful indications in order to 
improve the force models of the orbit determination software as far 
as tidal
perturbations are concerned and to perform new analyses with real
residuals. Moreover, when real data will be collected for a given
scenario it
will be possible to use them
in our
software  in order to adapt the simulation procedure to the new
situation; e.g. it is expected that the noise level in the near
future will diminish in view of improvements in laser ranging technology
and modeling.
Thus, we shall  repeat our analyses for $\D\m_{tides}$ when these new
results become available.

\noindent{\bf Acknowledgments\\}
\noindent The authors are grateful to I. Ciufolini for useful 
discussions. L. Iorio is  thoroughly indebted to N. Cufaro-Petroni 
for his
help and L.Guerriero for his support to him at Bari. E. C. Pavlis 
gratefully acknowledges partial support for this project
from NASA's Cooperative Agreement NCC 5-339.

\end{document}